\documentclass[12pt]{article}

\usepackage{amssymb}
\usepackage{amsmath}
\usepackage{latexsym}
\usepackage{yfonts}

\catcode`@=11
\renewcommand{\section}{\@startsection{section}{1}{0pt}{\medskipamount}
{\medskipamount}{\large\bf}}
\numberwithin{equation}{section}
\catcode`@=12

\def\beq{\begin{eqnarray}}    
\def\eeq{\end{eqnarray}}      

\def\pa{\partial}                       

\def\={\ =\ }

\begin{document}

\begin{center}

{\Large\bf Gauge (in)dependence and background field formalism}

\vspace{18mm}

{\large 
Peter M. Lavrov$^{(a, b)} \footnote{E-mail: lavrov@tspu.edu.ru}$,\;
}

\vspace{8mm}

\noindent  ${{}^{(a)}} ${\em
Tomsk State Pedagogical University,\\
Kievskaya St.\ 60, 634061 Tomsk, Russia}

\noindent  ${{}^{(b)}} ${\em
National Research Tomsk State  University,\\
Lenin Av.\ 36, 634050 Tomsk, Russia}

\vspace{20mm}

\begin{abstract}
\noindent
It is shown that the gauge invariance and gauge dependence properties of effective
action for Yang-Mills theories should be considered as two independent issues
in the background field formalism. Application of this formalism to formulate the
functional renormalization group approach is discussed. It is proven that
there is a possibility to construct the corresponding average effective action
invariant under the gauge transformations of background vector field. Nevertheless,
being gauge invariant this action  remains gauge dependent on-shell.
\end{abstract}

\end{center}

\vfill

\\

\newpage

\section{Introduction}
\noindent
Ii is well-known fact that
the gauge symmetry of an initial action is broken on quantum level because
of the gauge fixing procedure in process of quantization.
Generating functional of vertex functions
(effective action) being main quantity in quantum field theory depends on
gauges \cite{Jac,DJac,Niel,FK}. This dependence has a special form
and disappears on-shell  \cite{LT1,VLT}. In its turn it allows to have a
physical interpretation of results obtained on quantum level.

The background field method \cite{DeW, AFS,Abbott} presents
a reformulation of quantization procedure for Yang-Mills theories
allowing to work with  the effective action
invariant under the gauge transformations of background fields and
to reproduce all usual physical results by choosing a special background field
condition \cite{Abbott,Weinberg}.
Various aspects of quantum properties of gauge theories  have been successfully studied in
this technique \cite{'tH,K-SZ,GvanNW,CMacL,IO,GS,Ven,Gr,BC,FPQ}.
Application of the background field method  simplifies essentially
calculations of Feynman diagrams  in gauge theories
(among recent applications of this approach see, for example, \cite{BQ,Barv,FT, BLT-YM,BFMc}).
The gauge dependence problem in this method remains very important matter although
it  does not discuss  because standard considerations are restricted by
the background field gauge condition and by the invariance of generating functionals of Green functions under
gauge transformations of background fields.

In the present paper we study the gauge dependence of generating functionals
of Green functions in the background field formalism for Yang-Mills theories in class
of gauges depending on gauge and background vector fields.
The background field gauge condition belongs them as a special choice.
We prove that the gauge invariance can be achieved if the gauge fixing functions
satisfy a tensor transformation law and are linear in gauge fields. We consider
the gauge dependence and gauge invariance problems within the background
field formalism as two independent ones. To support this point of view we
analyze the functional renormalization group (FRG) approach \cite{Wet1,Wet2}
in the background field formalism. We find restrictions on tensor structure of
the regulator functions which allow to construct a gauge invariant average
effective action. Nevertheless, being gauge invariant this action remains
a gauge dependent quantity on-shell making impossible a physical interpretation
of results obtained for gauge theories.

The paper is organized as follows. Section 2 is devoted to description
of the background field formalism in gauges more general than the usual
background field gauge condition, to prove
the gauge independence of vacuum functional and to study symmetry properties of
the effective action. In Section 3
we analyze the gauge invariance of background average effective action for the
FRG approach and find restrictions on regulator functions
admitting this invariance. In section 4 we prove the gauge dependence of
vacuum functional (and therefore S-matrix)
for the FRG approach.
In section 5 concluding remarks are given.

In the paper the DeWitt's condensed notations are used \cite{DeWitt}.
We employ the notation $\varepsilon(A)$ for the
Grassmann parity and the ${\rm gh}(A)$ for the ghost number of any quantity $A$ .
All functional derivatives  are taken from the left.  The functional right
derivatives with respect to fields are marked by special
symbol $"\leftarrow"$.


\section{Background field formalism  for Yang-Mills theories}
\noindent
We start with a gauge theory of non-abelian vector fields
$A^{\alpha}_{\mu}(x)$
($\varepsilon(A^{\alpha}_{\mu}(x))=0,\; {\rm gh}(A^{\alpha}_{\mu}(x))=0$)
formulated in the
Minkowski space-time of arbitrary dimension with the action
\beq
\label{a1}
\mathcal{S}_{YM}(A)=\int dx\Big( -\frac{1}{4}G_{\mu \nu }^{\alpha
}(A(x))G_{\mu \nu }^{\alpha }(A(x))\Big) ,
\eeq
where the notation
\beq
\label{a2} G_{\mu \nu }^{\alpha }(A(x))=\pa_{\mu}A^{\alpha}_{\nu}(x)-
\pa_{\nu}A^{\alpha}_{\mu}(x) +gf^{\alpha \beta \gamma }A_{\mu }^{\beta
}(x)A_{\nu }^{\gamma }(x),
\eeq
is used. In relation (\ref{a2})
$f^{\alpha\beta\gamma}$ are structure coefficients of a compact
simple gauge Lie group and $g$ is a gauge interaction constant.
The action  (\ref{a1}) is invariant under gauge transformations with arbitrary
gauge functions  $\omega_{\alpha}(x)$,
 \beq
\delta_{\omega}\mathcal{S}_{YM}(A)=0,\quad
\delta _{\omega }A_{\mu }^{\alpha }(x)= \big( \partial _{\mu }\delta
_{\alpha \beta }+gf^{\alpha \sigma \beta }A_{\mu }^{\sigma }(x)\big)
\omega _{\beta }(x)=
D_{\mu }^{\alpha \beta }(A(x))\omega_{\beta }(x).
\label{a3}
\eeq

In the background field formalism \cite{DeW, AFS,Abbott}
the gauge field  $A^{\alpha}_{\mu}(x)$ appearing in classical action
(\ref{a1}), is replaced by  $A^{\alpha}_{\mu}(x)+{\cal B}^{\alpha}_{\mu}(x)$,
\beq
\label{a4}
\mathcal{S}_{YM}(A)\;\rightarrow \;\mathcal{S}_{YM}(A+{\cal B}),
\eeq
where ${\cal B}^{\alpha}_{\mu}(x)$ is considered as an external field.
The action $\mathcal{S}_{YM}(A+{\cal B})$ obeys obviously the gauge invariance,
\footnote{In what follows we will omit the space - time argument $x$ of fields and
gauge parameters
when this does not lead to misunderstandings
in the formulas and relations.}
\beq
\delta_{\omega}\mathcal{S}_{YM}(A+{\cal B})=0,\quad
\delta _{\omega }A_{\mu }^{\alpha }=
D_{\mu }^{\alpha \beta }(A+{\cal B})\omega_{\beta }.
\label{a5}
\eeq

The corresponding Faddeev-Popov action
$S_{FP}=S_{FP}(\phi,{\cal B})$
has the form \cite{FP}\footnote{The action (\ref{a6}) is written in so-called singular gauge fixing. Non-singular gauge fixing
corresponds to addition in the right-hand side of  (\ref{a6})  the term $\int dx B^{\alpha}g_{\alpha\beta}B^{\beta}$ where $g_{\alpha\beta}=g_{\beta\alpha}$ are elements of a constant invertible matrix.  The term  is invariant under BRST transformations  and does not spoil the renormalization properties of the theory under consideration.}
\beq
\label{a6}
S_{FP}=\mathcal{S}_{YM}(A+{\cal B})+\int dx
\Big[\overline{C}^{\alpha}\Big(\chi_{\alpha}(A,{\cal B})
\frac{\overleftarrow{\delta}}{\delta A^{\beta}_{\mu}}\Big)
D^{\beta\gamma}_{\mu}(A+{\cal B})C^{\gamma}+
B^{\alpha}\chi_{\alpha}(A,{\cal B})\Big],
\eeq
where $\chi_{\alpha}(A,{\cal B})$ are functions lifting the degeneracy
of the Yang-Mills action,
$\phi=\{\phi^i\}$  is the set of all fields
$\phi^i=(A^{\alpha}_{\mu},B^{\alpha}, C^{\alpha}, \overline{C}^{\alpha})$
($\varepsilon(\phi^i)=\varepsilon_i$)
with the Faddeev-Popov ghost and anti-ghost fields
$ C^{\alpha}, \overline{C}^{\alpha}$ ($\varepsilon(C^{\alpha})=
\varepsilon( \overline{C}^{\alpha})=1,\;
{\rm gh}(C^{\alpha})=-{\rm gh}(\overline{C}^{\alpha})=1$), respectively, and
the Nakanishi-Lautrup auxiliary fields $B^{\alpha}$
($\varepsilon(B^{\alpha})=0,\; {\rm gh}(B^{\alpha})=0$).
A standard choice of $\chi_{\alpha}(A,{\cal B})$
corresponding to the background field
gauge condition \cite{Abbott}, reads
\beq
\label{a7}
\chi_{\alpha}(A,{\cal B})=D^{\alpha\beta}_{\mu}({\cal B})A^{\beta}_{\mu}.
\eeq
In what follows the specific  form of  $\chi_{\alpha}(A,{\cal B})$ is not essential for all results obtained  but the property
of linearity of these functions  with respect to fields $A^{\alpha}_{\mu}$ plays a crucial role in the background-field formalism.

The action
 (\ref{a6}) is invariant under global supersymmetry (BRST symmetry) \cite{BRS1,T}
\beq
\label{a8}
\delta_{B} A^{\alpha}_{\mu}=D^{\alpha\beta}_{\mu}(A+{\cal B})C^{\beta}\mu,\quad
\delta_{B} C^{\alpha}=\frac{g}{2}
f^{\alpha \beta \gamma }C^{\beta }C^{\gamma }\mu,\quad
\delta_{B}\overline{C}^{\alpha}=B^{\alpha} \mu,\quad
\delta_{B} B^{\alpha} =0,
\eeq
where  $\mu$ is a constant anti-commuting parameter or, in short,
\beq
\label{a9}
\delta_{B}\phi^i=R^i(\phi, {\cal B})\mu,\quad
\varepsilon(R^i(\phi, {\cal B}))=\varepsilon_i+1,
\eeq
where
\beq
\label{a10}
R^i(\phi, {\cal B})=\big(D^{\alpha\beta}_{\mu}(A+{\cal B})C^{\beta},\; 0\;, \frac{g}{2}
f^{\alpha \beta \gamma }C^{\beta }C^{\gamma }, B^{\alpha}\big).
\eeq
Introducing the gauge fixing functional $\Psi=\Psi(\phi,{\cal B})$,
\beq
\label{a11}
\Psi=\int dx \;\overline{C}^{\alpha}\chi_{\alpha}(A,{\cal B}),
\eeq
the action (\ref{a6}) rewrites in the form
\beq
\label{a12}
S_{FP}(\phi,{\cal B})=\mathcal{S}_{YM}(A+{\cal B})+\Psi(\phi,{\cal B})
{\hat R}(\phi,{\cal B}),
\qquad \mathcal{S}_{YM}(A+{\cal B}){\hat R}(\phi,{\cal B})=0,
\eeq
where
\beq
\label{a13}
{\hat R}(\phi,{\cal B})=\int dx\;
\frac{\overleftarrow{\delta}}{\delta\phi^i}R^i(\phi, {\cal B})
\eeq
is the generator of BRST transformations.
Due to the nilpotency property of ${\hat R}$, ${\hat R}^2=0$, the BRST symmetry of $S_{FP}$
follows from the presentation (\ref{a12}) immediately,
\beq
\label{a14}
S_{FP}(\phi,{\cal B}){\hat R}(\phi,{\cal B})=0.
\eeq

The generating functional of Green functions in the background field
method is defined in the form of functional integral
\beq
\label{a15}
Z(J, {\cal B})=\int
d\phi\;\exp\Big\{\frac{i}{\hbar}\big[S_{FP}(\phi, {\cal
B})+J\phi\big]\Big\}=\exp\Big\{\frac{i}{\hbar}W(J, {\cal B})\Big\},
\eeq where $W(J, {\cal B})$ is the generating functional of
connected Green functions. In (\ref{a15}) the notations
\beq
\label{a16}
J\phi=\int dx J_i(x)\phi^i(x), \quad
J_i(x)=(J^{\alpha}_{\mu}(x), J^{(B)}_{\alpha}(x), \overline{J}_{\alpha}(x), J_{\alpha})(x)
\eeq
are used and
$J_i(x)$ \big($\varepsilon(J_i(x))=\varepsilon_i,\;{\rm gh}(J_i(x))={\rm gh}(\phi^i(x))$\big)
are external sources to fields $\phi^i(x)$.

Let $Z_{\Psi}({\cal B})$ be the vacuum functional which corresponds
to the choice of gauge fixing functional (\ref{a11}) in the presence
of external fields ${\cal B}$, \beq \label{a17} Z_{\Psi}({\cal
B})=\int
d\phi\;\exp\Big\{\frac{i}{\hbar}\big[\mathcal{S}_{YM}(A+{\cal
B})+\Psi(\phi,{\cal B}) {\hat R}(\phi,{\cal B})\big]\Big\}=\int
d\phi\;\exp\Big\{\frac{i}{\hbar}S_{FP}(\phi, {\cal B})\Big\}. \eeq
In turn, let $Z_{\Psi+\delta\Psi}$ be the vacuum functional
corresponding to a gauge fixing functional $\Psi(\phi,{\cal
B})+\delta\Psi(\phi,{\cal B})$,
\beq
\label{a18}
Z_{\Psi+\delta\Psi}({\cal B})=\int
d\phi\;\exp\Big\{\frac{i}{\hbar}\big[S_{FP}(\phi, {\cal
B})+\delta\Psi(\phi,{\cal B}){\hat R}(\phi,{\cal B})\big]\Big\}.
\eeq
Here, $\delta\Psi(\phi,{\cal B})$
is an arbitrary infinitesimal odd functional
which  may,  in general,  have  a form  differing on (\ref{a11}).
Making use of the change of variables $\phi^i$ in the form of BRST
transformations (\ref{a9}) but with replacement of the constant parameter
$\mu$ by the following functional
\beq
\label{a19}
\mu=\mu(\phi,{\cal B})=\frac{i}{\hbar}\delta\Psi(\phi,{\cal B}),
\eeq
and taking into account that the Jacobian of transformations is
equal to
\beq
\label{a20} J=\exp\{-\mu(\phi,{\cal B}){\hat
R}(\phi,{\cal B})\},
\eeq
we find the gauge independence of the
vacuum functional
\beq
\label{a21}
Z_{\Psi}({\cal
B})=Z_{\Psi+\delta\Psi}({\cal B}).
\eeq
The property (\ref{a21}) was
a reason to omit the label $\Psi$ in the definition of generating
functionals (\ref{a15}). In deriving (\ref{a21}) the relation
\beq
\label{a25}
 (-1)^{\varepsilon_i}\frac{\pa}{\pa\phi^i}R^i(\phi, {\cal B})=0,
\eeq
was used. It holds due to the antisymmetry property of structure constants,
$f^{\alpha\beta\gamma}=-f^{\beta\alpha\gamma}$. In turn, the property (\ref{a21})
means that due to the equivalence theorem \cite{KT}
the physical $S$-matrix does not depend on the gauge fixing.

The vacuum functional $Z({\cal B})=Z(J=0, {\cal B})$ obeys the very
important property of gauge
invariance with respect to gauge transformations of external fields,
\beq
\label{a26}
\delta_{\omega}{\cal B}_{\mu}^{\alpha}=D_{\mu}^{\alpha\beta}({\cal B})\omega_{\beta},\quad
\delta_{\omega}Z({\cal B})=0.
\eeq
It means the gauge invariance of functional $W({\cal B})=W(J=0, {\cal B})$,
$\delta_{\omega}W({\cal B})=0$, as well. The proof is based on using the
change of variables in
the functional integral (\ref{a17}) of the following form
\beq
\nonumber
&&\delta_{\omega}A_{\mu}^{\alpha}=g f^{\alpha\gamma\beta}A_{\mu}^{\gamma}\omega_{\beta},
\quad \delta_{\omega}C^{\alpha}=gf^{\alpha\gamma\beta}C^{\gamma}\omega_{\beta},\\
\label{a27}
&& \delta_{\omega} \overline{C}^{\alpha}=
g f^{\alpha\gamma\beta} \overline{C}^{\gamma}\omega_{\beta},\quad  \delta_{\omega}B^{\alpha}=
gf^{\alpha\gamma\beta}B^{\gamma}\omega_{\beta}
\eeq
taking into account that the Jacobian of transformations (\ref{a27}) is equal to a unit,
and assuming the transformation law of gauge fixing functions $\chi_{\alpha}$ according to
\beq
\label{a28}
\delta_{\omega}\chi_{\alpha}(A, {\cal B})=
g f^{\alpha\gamma\beta}\chi_{\gamma}(A, {\cal B})\omega_{\beta},
\eeq
which is fulfilled explicitly for the background field gauge condition (\ref{a7}). In particular,
it can be argued the invariance of $S_{FP}(\phi, {\cal B})$ under combined gauge
transformations
(\ref{a26}) and (\ref{a27})
\beq
\label{a29}
\delta_{\omega}S_{FP}(\phi, {\cal B})=0.
\eeq

 The Slavnov-Taylor identity for the generating functional of Green functions is derived
 in standard manner,
\beq
\label{a30}
\int dx J_iR^i\Big(\frac{\hbar}{i}\frac{\delta}{\delta J},{\cal B}\Big)Z(J,{\cal B})=0,
\eeq
as consequence of the BRST symmetry of $S_{FP}$ (\ref{a14}) on the quantum level.
In terms of generating functional of connected Green functions, $W(J, {\cal B})$,
the identity (\ref{a30}) rewrites as
\beq
\label{a31}
\int dx J_iR^i\Big(\frac{\delta W(J, {\cal B})}{\delta J}+
\frac{\hbar}{i}\frac{\delta}{\delta J},{\cal B}\Big)\cdot 1=0.
\eeq

The generating functional of vertex functions (effective action),
$\Gamma=\Gamma(\Phi, {\cal B})$,
is defined in a standard form through the Legendre transformation of $W(J, {\cal B})$,
\beq
\label{a32}
\Gamma(\Phi, {\cal B})= W(J, {\cal B})-\int dx J_i\Phi^i,\quad
\Phi^i=\frac{\delta{W}(J)}{\delta J_i},\quad
\Phi^i=({\cal A}^{\alpha}_{\mu}, \Phi^{\alpha}_{(B)}, {\cal C}^{\alpha},
\overline{{\cal C}}^{\alpha}),
\eeq
so that
\beq
\label{a33}
\Gamma(\Phi, {\cal B})\frac{\overleftarrow{\delta}}{\delta\Phi^i} =-J_i.
\eeq
The Ward identity (\ref{a31}) rewrites for $\Gamma(\Phi, {\cal B})$ in the form
\beq
\label{a34}
\Gamma(\Phi, {\cal B}) \widehat{{\overline{R}}}(\Phi, {\cal B})=0,
\eeq
where
\beq
\label{a35}
\widehat{{\overline{R}}}(\Phi, {\cal B})= \int dx \frac{\overleftarrow{\delta}}{\delta\Phi^i}
\overline{R}^i(\Phi, {\cal B}), \quad
\overline{R}^i(\Phi, {\cal B})=R^i({\hat \Phi}, {\cal B})\cdot 1,
\eeq
can be considered as the generator of quantum BRST transformations.
In relation (\ref{a35}) the notations
\beq
\label{a36}
{\hat \Phi}^i(x)=\Phi^i(x)+i\hbar\int dy
(\Gamma^{'' -1})^{ij}(\Phi, {\cal B})(x,y)\frac{\overrightarrow{\delta}}{\delta\Phi^j(y)},
\eeq
are used. In turn the matrix
$(\Gamma^{'' -1})^{ij}(x,y)=(\Gamma^{'' -1})^{ij}(\Phi, {\cal B})(x,y)$
is inverse to the matrix of second derivatives of effective action,
\beq
\label{a37}
&&(\Gamma^{''})_{ij}(\Phi, {\cal B})(x,y)=
\frac{\overrightarrow{\delta}}{\delta\Phi^i(x)}\Big(\Gamma(\Phi, {\cal B})
\frac{\overleftarrow{\delta}}{\delta\Phi^j(y)}\Big),\\
&& \int dz (\Gamma^{''
-1})^{ik}(x,z)(\Gamma^{''})_{kj}(z,y)=\delta^i_j \delta(x-y).
\eeq
The Ward identity (\ref{a34}) can be interpreted as the invariance
of effective action $\Gamma(\Phi, {\cal B})$ under the {\it quantum}
BRST transformations of $\Phi^i$ with generators ${\bar R}^i(\Phi,
{\cal B})$.

Notice that in the case of anomaly-free theories and a
regularization preserving the gauge invariance,  one can prove
in the standard manner \cite{VLT}
(see also \cite{Weinberg}) that the renormalized action
$S_{FP, ren}(\phi,{\cal B})$ and
the renormalized effective action $\Gamma_{ren}(\Phi,{\cal B})$ satisfy
the same equations
(\ref{a14}) and (\ref{a34}) with the corresponding nilpotent operators
${\hat R}_{ren}(\phi,{\cal B})$ and
$\widehat{{\overline{R}}}_{ren}(\Phi, {\cal B})$, respectively.

The invariance of $S_{FP}$ (\ref{a29}) means that the functional $Z(J, {\cal B})$
is invariant
\beq
\label{a38}
\!\!Z(J, {\cal B})\!\!\int \!\!dx \frac{\overleftarrow{\delta}}{\delta{\cal B}^{\alpha}_{\mu}}
D^{\alpha\beta}_{\mu}({\cal B})\omega_{\beta}=
gf^{\alpha\gamma\beta}\omega_{\beta}\!\!\int \!\!dx
\Big(\!J^{\alpha}_{\mu}\frac{\delta}{\delta J^{\gamma}_{\mu}}\!+\!
J_{\alpha}\frac{\delta}{\delta J_{\gamma}}\!+\!
\overline{J}_{\alpha}\frac{\delta}{\delta \overline{J}_{\gamma}}\!+\!
J^{(B)}_{\alpha}\frac{\delta}{\delta J^{(B)}_{\gamma}}\Big)\!Z(J, {\cal B}),
\eeq
under the gauge transformations of the background vector field ${\cal B}$ (\ref{a26})
and simultaneously the tensor
transformations of sources
\beq
\label{a39}
\delta_{\omega}J^{\alpha}_{\mu}=g f^{\alpha\gamma\beta}J^{\gamma}_{\mu}\omega_{\beta},\;
\delta_{\omega}\overline{J}_{\alpha}=
g f^{\alpha\gamma\beta}\overline{J}_{\gamma}\omega_{\beta},\;
\delta_{\omega}J_{\alpha}=g f^{\alpha\gamma\beta}J_{\gamma}\omega_{\beta},\;
\delta_{\omega}J^{(B)}_{\alpha}=g f^{\alpha\gamma\beta}J^{(B)}_{\gamma}\omega_{\beta}.
\eeq
In its turn the functional $W(J, {\cal B})$ obeys the same symmetry property as well,
\beq
\label{a40}
\!\!\!W(J, {\cal B})\!\!\int\! dx \frac{\overleftarrow{\delta}}{\delta{\cal B}^{\alpha}_{\mu}}
D^{\alpha\beta}_{\mu}({\cal B})\omega_{\beta}\!=\!
gf^{\alpha\gamma\beta}\omega_{\beta}\!\!\int \!\!dx
\Big(\!J^{\alpha}_{\mu}\frac{\delta}{\delta J^{\gamma}_{\mu}}\!+\!
J_{\alpha}\frac{\delta}{\delta J_{\gamma}}\!+\!
\overline{J}_{\alpha}\frac{\delta}{\delta \overline{J}_{\gamma}}\!+\!
J^{(B)}_{\alpha}\!\frac{\delta}{\delta J^{(B)}_{\gamma}}\Big)\!W(J, {\cal B}).
\eeq
In terms of the functional $\Gamma(\Phi, {\cal B})$ the relation (\ref{a40}) reads
\beq
\label{a41}
\!\!\!\Gamma(\Phi, {\cal B})\!\!\!\int \!\!dx \!\frac{\overleftarrow{\delta}}{\delta{\cal B}^{\alpha}_{\mu}}
D^{\alpha\beta}_{\mu}({\cal B})\omega_{\beta}\!=\!-
\Gamma(\Phi, {\cal B})\!\!\!\int\!\! dx \Big(\!\frac{\overleftarrow{\delta}}{\delta{\cal A}^{\alpha}_{\mu}}
{\cal A}^{\gamma}_{\mu}\!\!+\!\frac{\overleftarrow{\delta}}{\delta{\cal C}^{\alpha}}
{\cal C}^{\gamma}\!\!+\!\frac{\overleftarrow{\delta}}{\delta\overline{{\cal C}}^{\alpha}}
\overline{{\cal C}}^{\gamma}\!\!+\!\frac{\overleftarrow{\delta}}{\delta{\Phi^{\alpha}_{(B)}}}
{\Phi^{\gamma}_{(B)}}
\Big)g f^{\alpha\gamma\beta}\omega_{\beta}.
\eeq
The relation (\ref{a41}) proves the invariance of $\Gamma(\Phi, {\cal B})$ under the gauge
transformation of external vector field ${\cal B}$ accompanied by the tensor transformations
of fields ${\cal A}, {\cal C}, \overline{{\cal C}}, \Phi_{(B)}$,
\beq
\label{a42}
\delta_{\omega}{\cal A}^{\alpha}_{\mu}=gf^{\alpha\gamma\beta}
{\cal A}^{\gamma}_{\mu}\omega_{\beta},\;
\delta_{\omega}{\cal C}^{\alpha}=gf^{\alpha\gamma\beta}{\cal C}^{\gamma}\omega_{\beta},\;
\delta_{\omega}\overline{{\cal C}}^{\alpha}=gf^{\alpha\gamma\beta}
\overline{{\cal C}}^{\gamma}\omega_{\beta},\; \delta_{\omega}\Phi^{\alpha}_{(B)}=
gf^{\alpha\gamma\beta} \Phi_{(B)}^{\gamma}\omega_{\beta}.
\eeq
From (\ref{a41}) it follows the main property of functional
 $\Gamma({\cal B})=\Gamma(\Phi,{\cal B})|_{\Phi=0}$ in the background field formalism
 \footnote{In the present paper we do not discuss a role of the BRST- and   background gauge symmetries and problems connected with renormalization programm for gauge theories  within the background field method refereeing to the papers \cite{Gr,BC,FPQ}.}
 \beq
\label{a43}
\Gamma({\cal B})\int dx \frac{\overleftarrow{\delta}}{\delta{\cal B}^{\alpha}_{\mu}}
D^{\alpha\beta}_{\mu}({\cal B})\omega_{\beta}=0.
\eeq
The relations between the standard generating functionals and the analogous quantities
in the background field formalism are established with modification of gauge functions
likes to
$\chi_{\alpha}(A,{\cal B})\rightarrow \chi^{'}_{\alpha}(A,{\cal B})=\chi_{\alpha}(A,{\cal B})-\pa_{\mu}{\cal B}^{\alpha}_{\mu}$ \cite{Abbott}.

\section{Gauge invariance of average effective action}
\noindent
In this section we discuss the gauge invariance of average effective action
for the FRG  \cite{Wet1,Wet2} in the
background field formalism. Of course this issue is not new (see,
for example, \cite{FLP,Wet3}), but we are going to demonstrate that requirement of
gauge invariance of the average effective action restricts a tensor structure of
regulator functions being essential objects of the approach.
One of main ideas of the functional renormalization group approach
was to modify behavior of propagators of vector and ghost fields in
IR and UV regions with the help of addition of a scale-dependent  regulator action
being quadratic in the fields.
The scale-dependent regulator action
\beq
\label{b1}
S_k(\phi)=\int dx \Big[\frac{1}{2}A^{\alpha}_{\mu}(x)
R^{(1)\mu\nu}_{k\;\alpha\beta}(x)A^{\beta}_{\nu}(x)
+
\overline{C}^{\alpha}(x)R^{(2)}_{k\;\alpha\beta}(x)C^{\beta}(x)\Big]
\eeq
is defined by regulator functions
$R^{(1)\mu\nu}_{k\;\alpha\beta}(x), R^{(2)}_{k\;\alpha\beta}(x)$ which are independent
of fields. The regulator functions $R^{(1)\mu\nu}_{k\;\alpha\beta}$ obey evident symmetry
properties
\beq
\label{b2}
R^{(1)\mu\nu}_{k\;\alpha\beta}=R^{(1)\nu\mu}_{k\;\beta\alpha}.
\eeq

Let us require the invariance of $S_k(\phi)$ under transformations (\ref{a21})
\beq
\label{b3}
\delta_{\omega}S_k(\phi)=0.
\eeq
It leads to the equations
\beq
\label{b4}
f^{\alpha\beta\sigma}R^{(1)\mu\nu}_{k\;\sigma\gamma} +
R^{(1)\mu\nu}_{k\;\alpha\sigma}f^{\sigma\gamma\beta}=0,
\quad
f^{\alpha\beta\sigma}R^{(2)}_{k\;\sigma\gamma} +
R^{(2)}_{k\;\alpha\sigma}f^{\sigma\gamma\beta}=0,
\eeq
which can be presented in terms of Lie group generators
$(t^{\alpha})_{\beta\gamma}=f^{\beta\alpha\gamma}$ as
\beq
\label{b5}
[t^{\beta}, R^{(1)\mu\nu}_k]_{\alpha\gamma}=0,\quad
[t^{\beta}, R^{(2)}_k]_{\alpha\gamma}=0.
\eeq
Due to the Schur's lemma it follows from (\ref{b5}) that
\beq
\label{b6}
R^{(1)\mu\nu}_{k\;\alpha\beta}=\delta_{\alpha\beta}R^{(1)\mu\nu}_{k},\quad
R^{(2)}_{k\;\alpha\beta}=\delta_{\alpha\beta}R^{(2)}_{k},
\eeq
Therefore the regulator action (\ref{b1}) should be of the form
\beq
\label{b7}
S_k(\phi)=\int dx \Big[\frac{1}{2}A^{\alpha}_{\mu}(x)R^{(1)\mu\nu}_{k}(x)
A^{\alpha}_{\nu}(x)+
\overline{C}^{\alpha}(x)R^{(2)}_k(x)C^{\alpha}(x)\Big]
\eeq
to retain the invariance (\ref{b3}). In this case
 the full action
\beq
\label{b8}
S_{k}(\phi, {\cal B})=S_{FP}(\phi, {\cal B})+ S_k(\phi),
\eeq
is
invariant under transformations (\ref{a21}),
\beq
\label{b9}
\delta_{\omega}S_{k}(\phi, {\cal B})=0.
\eeq
The invariance (\ref{b9}) allows to extend all previous result concerning the
gauge invariance problem on quantum level.
The generating functionals of Green functions $Z_k(J, {\cal B})$
and connected Green functions
$W_k(J, {\cal B})$  are defined by the
functional integral
\beq
\label{b10}
Z_k(J, {\cal B})=\int
d\phi\;\exp\Big\{\frac{i}{\hbar}\big[S_{FP}(\phi, {\cal
B})+S_k(\phi)+J\phi\big]\Big\}=\exp\Big\{\frac{i}{\hbar}W_k(J, {\cal B})\Big\},
\eeq
Repeating the same arguments as in previous section,  we can proof
the gauge invariance of the vacuum functional
$Z_k({\cal B})=Z_k(0, {\cal B})$ for the FRG  approach in
the background field formalism
\beq
\label{b11}
\delta_{\omega}Z_k({\cal B})=0,\quad \delta_{\omega}{\cal B}^{\alpha}_{\mu}=
D_{\mu}^{\alpha\beta}({\cal B})\omega_{\beta}.
\eeq
From (\ref{b9}) and (\ref{b10}) it follows the gauge invariance of functional
$W_k({\cal B})=W_k(0, {\cal B})$ as well,
\beq
\label{b12}
\delta_{\omega}W_k({\cal B})=0.
\eeq
In similar way we can proof the gauge invariance of average effective action
$\Gamma_k(\Phi, {\cal B})=W_k(J, {\cal B})-J\Phi$,
\beq
\label{b13}
\!\!\Gamma_k(\Phi, {\cal B})\frac{\overleftarrow{\delta}}{\delta{\cal B}^{\alpha}_{\mu}}
D^{\alpha\beta}_{\mu}({\cal B})\omega_{\beta}=-
\Gamma_k(\Phi, {\cal B})\Big(\frac{\overleftarrow{\delta}}{\delta{\cal A}^{\alpha}_{\mu}}
{\cal A}^{\gamma}_{\mu}+\frac{\overleftarrow{\delta}}{\delta{\cal C}^{\alpha}}
{\cal C}^{\gamma}+\frac{\overleftarrow{\delta}}{\delta\overline{{\cal C}}^{\alpha}}
\overline{{\cal C}}^{\gamma}+\frac{\overleftarrow{\delta}}{\delta{\Phi^{\alpha}_{(B)}}}
{\Phi^{\gamma}_{(B)}}\Big)g f^{\alpha\gamma\beta}\omega_{\beta}
\eeq
because the derivation of (\ref{b13}) operates in fact with the invariance of full action,
$\delta_{\omega}(S_{FP}(\phi, {\cal B})+S_k(\phi))=0$, only. In particular,
it follows from (\ref{a12}) the statement
\beq
\label{b14}
\Gamma_k({\cal B})\frac{\overleftarrow{\delta}}{\delta{\cal B}^{\alpha}_{\mu}}
D^{\alpha\beta}_{\mu}({\cal B})\omega_{\beta}=0, \quad
\Gamma_k({\cal B})=\Gamma_k(\Phi, {\cal B})|_{\Phi=0},
\eeq
concerning the invariance of $\Gamma_k({\cal B})$ under
the gauge transformations of external vector field.

\section{Gauge dependence  of average effective action}
\noindent
In this section we are going to investigate the gauge dependence problem for the
FRG  approach  in the background field formalism.
Standard formulation of this method being applied to gauge theories leads to ill defined the
average effective action and the corresponding flow equation which still remain
gauge dependent even on-shell \cite{LSh,LM}. The last feature of the FRG
approach does not give a possibility of physical interpretations of results obtained.

To support our understanding the independence of gauge invariance and gauge dependence
problems within background field formalism let us consider the generating functionals of
Green functions  and connected Green functions supplied   with label $"\Psi"$
\beq
\nonumber
Z_{k\Psi}(J, {\cal B})&=&\int
d\phi\;\exp\Big\{\frac{i}{\hbar}\big[S_{YM}( {\cal A}+ {\cal B})+
\Psi(\phi,{\cal B}){\hat R}(\phi, {\cal B})
+S_k(\phi)+J\phi\big]\Big\}=\\
\label{c1}
&=&\int
d\phi\;\exp\Big\{\frac{i}{\hbar}S_{k}(\phi, {\cal B})
\Big\}= \exp\Big\{\frac{i}{\hbar}W_{k\Psi}(J, {\cal B})\Big\},
\eeq
Taking into account that the regulator action does not depend
on gauge we consider the functional  (\ref{c1}) at $J=0$ corresponding another choice
of the gauge fixing functional $\Psi \rightarrow \Psi+\delta\Psi$
\beq
\label{c2}
Z_{k\Psi+\delta\Psi}({\cal B})=\int
d\phi\;\exp\Big\{\frac{i}{\hbar}\big[S_{k}(\phi, {\cal B})+
\delta\Psi(\phi,{\cal B}){\hat R}(\phi, {\cal B})
\big]\Big\}=\exp\Big\{\frac{i}{\hbar}W_{k\Psi+\delta\Psi}({\cal B})\Big\},
\eeq
where
\beq
\label{c3}
\delta\Psi=\delta\Psi(\phi, {\cal B})=\int dx \overline{C}^{\alpha}
\delta\chi_{\alpha}(A, {\cal B}).
\eeq

Now we are trying to compensate additional term $\delta\Psi{\hat R}$
in the exponent (\ref{c2}) using the changes of variables in the functional integral
related closely to the symmetry of actions  $S_{FP}(\phi, {\cal B})$ (\ref{a14}) and
$S_k(\phi, {\cal B})$ (\ref{b8}). In the functional integral (\ref{c2}) we make first a
change of variables in the form of the BRST transformations (\ref{a9}), (\ref{a10}),
 but trading the constant parameter $\mu$ to a functional $\Lambda=\Lambda(\phi, {\cal B})$.
 The action $S_{FP}$ (\ref{a12}) is invariant under
 such change of variables but the action $S_k(\phi)$ (\ref{b7}) is not invariant,
 with the  following variation
 \beq
\label{c4}
\delta S_k(\phi)=\int dx\Big(A^{\alpha}_{\mu}R^{(1)\mu\nu}D^{\alpha\beta}_{\nu}
(A+{\cal B})C^{\beta}+\frac{1}{2}\overline{C}^{\alpha}R^{(2)}_k
f^{\alpha\beta\gamma}C^{\beta}C^{\gamma}
-B^{\alpha}R^{(2)}_k C^{\alpha}\Big)\Lambda .
\eeq
The corresponding Jacobian $J_1$ reads
\beq
\label{c5}
J_1=\exp\Big\{-\int dx
\Big(\frac{\delta \Lambda}{\delta A^{\alpha}_{\mu}}D^{\alpha\beta}_{\mu}
(A+{\cal B})C^{\beta}+\frac{1}{2}f^{\alpha\beta\gamma}C^{\beta}C^{\gamma}
\frac{\delta\Lambda}{\delta C^{\alpha}}
+\frac{\delta\Lambda}{\delta\overline{C}^{\alpha}}B^{\alpha}\Big)\Big\}.
\eeq
We make additionally a change of variables related to gauge transformations
(\ref{a26}), (\ref{a27}) but using instead of parameters $\omega_{\alpha}(x)$ functions
$\Omega_{\alpha}(x)=\Omega_{\alpha}(x,\phi(x), {\cal B}(x))$.
The action $S_k(\phi, {\cal B})$
is invariant
under these transformations but the relevant Jacobian, $J_2$ is not trivial,
\beq
\label{c6}
J_2=\exp\Big\{gf^{\alpha\beta\gamma}
\int dx\Big(A^{\beta}_{\mu}(x)\frac{\pa\Omega_{\gamma}(x)}{\pa A^{\alpha}_{\mu}(x)}
-C^{\beta}(x)\frac{\pa\Omega_{\gamma}(x)}{\pa C^{\alpha}(x)}-
\overline{C}^{\beta}(x)
\frac{\pa\Omega_{\gamma}(x)}{\pa \overline{C}^{\alpha}(x)}\Big)\Big\}.
\eeq
If the condition,
\beq
\label{c7}
J_1J_2\exp\Big\{\frac{i}{\hbar}\int dx \big[
\delta\Psi(\phi,{\cal B}){\hat R}(\phi, {\cal B})
+\delta S_k(\phi)]\Big\}=1,
\eeq
is satisfied then the functional $Z_{k \Psi}({\cal B})$  does not
depend on gauge fixing functional $\Psi$. Having in mind the ghost numbers and Grassmann
parities of functional $\Lambda$ and functions $\Omega_{\alpha}(x)$
\beq
\label{c8}
{\rm gh}(\Lambda)=-1, \quad {\rm gh}(\Omega_{\alpha}(x))=0, \quad
\varepsilon(\Lambda)=1,\quad \varepsilon(\Omega_{\alpha}(x))=0,
\eeq
we have the following presentation in the lower power of ghost fields,
\beq
\label{c9}
\Lambda=\Lambda^{(1)}+\Lambda^{(3)},\quad
\Omega_{\alpha}(x)=\Omega_{\alpha}^{(0)}(x)+\Omega_{\alpha}^{(2)}(x),
\eeq
where
\beq
\label{c10}
&&\Lambda^{(1)}=\int dx \overline{C}^{\alpha}(x)
\lambda^{(1)}_{\alpha}(x, A(x), {\cal B}(x)),\\
&&\Lambda^{(3)}=\int dx \frac{1}{2}
\overline{C}^{\alpha}(x)\overline{C}^{\beta}(x)
\lambda^{(3)}_{\alpha\beta\gamma}(x,A(x),{\cal B}(x))C^{\gamma}(x),\\
\label{c11}
&& \Omega_{\alpha}^{(0)}(x)=\Omega_{\alpha}^{(0)}(x,A(x),{\cal B}(x)),\\
&&\Omega_{\alpha}^{(2)}(x, A(x),{\cal B}(x))= \overline{C}^{\beta}(x)
\omega^{(2)}_{\alpha\beta\gamma}(x, A(x), {\cal B}(x))C^{\gamma}(x).
\eeq
Vanishing terms in (\ref{c7}) which don't depend on ghost fields $C, \overline{C}$
and  auxiliary field $B$
leads to the condition
\beq
\label{c12}
\Omega_{\alpha}^{(0)}(x, A(x),{\cal B}(x))=0.
\eeq
Consider in the equation (\ref{c7}) terms linear in $B$ then we obtain
\beq
\label{c13}
\lambda^{(1)}_{\alpha}(x, A(x), {\cal B}(x))=
\frac{i}{\hbar}\delta\chi_{\alpha}(x,A(x),{\cal B}(x)).
\eeq
In turn analyzing the structures $B\overline{C}C$ in (\ref{c7}) we find the expression for
$\lambda^{(3)}_{\alpha\beta\gamma}$,
\beq
\label{c14}
&&\lambda^{(3)}_{\alpha\beta\gamma}(x,A,{\cal B})=
R^{(2)}(x)\big(\delta_{\beta\gamma}\lambda^{(1)}_{\alpha}(A,{\cal B})-\delta_{\alpha\gamma}
\lambda^{(1)}_{\beta}(A,{\cal B})\big),\\
\label{c15}
&&
\qquad\qquad\lambda^{(1)}_{\alpha}(A,{\cal B})=
\int dx \lambda^{(1)}_{\alpha}(x,A(x),{\cal B}(x)).
\eeq
Vanishing structures $\overline{C}C$ leads to algebraic equations
for $\omega^{(2)}_{\alpha\beta\gamma}$,
\beq
\nonumber
&&f^{\gamma\alpha\sigma}\omega^{(2)}_{\sigma\beta\gamma}(x, A(x), {\cal B}(x))+
f^{\gamma\beta\sigma}\omega^{(2)}_{\sigma\gamma\alpha}(x, A(x), {\cal B}(x))=\\
\label{c16}
&&
=\frac{i}{g\hbar}D^{\gamma\alpha}_{\nu}(A+{\cal B})
\big(A^{\gamma}_{\mu}(x)R^{(1)\mu\nu}_k(x)\big)\lambda^{(1)}_{\beta}(A,{\cal B}).
\eeq
Therefore, in the case (\ref{c9})-(\ref{c16}) we can reduce to zero in (\ref{c7})
all terms of the lowest order in fields $C,\overline{C}, B$.
Unfortunately, in its turn the $\lambda^{(3)}_{\alpha\beta\gamma}$ (\ref{c14}) creates
the non-local term of structure $B\overline{C}\;\! \overline{C}\;\! C \;\!C$ which
cannot be eliminated in a proposed scheme. It is necessary to add  for functional
$\Lambda$ and functions $\Omega_{\alpha}$ new terms of higher orders in ghost fields
up to infinity.
This situation looks  unsatisfactory in terms of conventional quantum field theory
and we
are forced to restrict ourself by the case when $\Omega_{\alpha}=0$ and
$\Lambda=\Lambda^{(1)}$.
Then we have
\beq
\label{c17}
Z_{k\Psi+\delta\Psi}({\cal B})=\int
d\phi\;\exp\Big\{\frac{i}{\hbar}\big[S_{k}(\phi, {\cal B})+
\delta S_k(\phi)\big]\Big\},
\quad Z_{k\Psi}({\cal B})\neq Z_{k\Psi+\delta\Psi}({\cal B}).
\eeq
Vacuum functional  in the FRG  approach
within  the background field formalism  remains gauge dependent similar
to  the standard formulation \cite{LSh,LM}. The same statement is valid
for elements of S-matrix due to
the equivalence theorem \cite{KT}. There are no problems deriving a modified Ward identity
which is a consequence of BRST invariance of action $S_{FP}(\phi, {\cal B})$ and
identities which follow from gauge
invariance of the action $S_k(\phi,{\cal B})$ as well as to study gauge dependence
of average effective action on-shell. We omit all these issues of the FRG
approach because they do not help to solve the gauge dependence
problem of results which are obtained within this method.

\section{Summary}
\noindent
In the present paper we have analyzed the problems of the gauge invariance and
gauge dependence of the generating functionals of Green functions in the
background field formalism. It should be stressed that
the gauge invariance of  background effective action is usually under intensive study because
it is a very important property for real calculations of Feynman diagrams.
In turn the gauge dependence problem remains not in a focus of studies within this formalism
although by itself this problem plays a principal role in our understanding of the ability
to give a consistent physical interpretation of quantum results for gauge theories.
We have supported this point of view by analysing the FRG approach
in the background field formalism. We have shown that although
the gauge invariance can be achieved with restrictions on the tensor structure of regulator
functions but the gauge dependence problem cannot be solved  in
the existing representation of the FRG approach
for gauge theories. The reason for this is the existing choice of regulator action (\ref{b7}).
Consistent quantization of gauge theories permits modifications of quantum action
($S_{FP}$ in the case of Yang-Mills theories) with the BRST-invariant additions
only \cite{BL}. The regulator action (\ref{b7}) is not BRST-invariant
that caused the gauge dependence problem.


\section*{Acknowledgments}

 The author thanks I.V. Tyutin for useful discussions. The work is supported in part by the
Ministry of Education and Science of the Russian Federation, grant
3.1386.2017 and by the RFBR grant 18-02-00153.

\end{document}